\documentclass[amsmat,amssymb,amsfonts,aps,prb,twocolumn,showpacs]{revtex4}

\usepackage{graphicx}
\usepackage{dcolumn}
\usepackage{bm}
\begin{document}

\title{Transport in magnetically ordered Pt nanocontacts}

\author{J. Fern\'andez-Rossier, D. Jacob, C. Untiedt, and J. J. Palacios}
\affiliation{Departamento de F\'{\i}sica Aplicada and Instituto Universitario de Materiales de Alicante (IUMA), Universidad de Alicante, San Vicente del Raspeig, 03690 Spain.}

\date{\today}
 
\begin{abstract}

Pt nanocontacts, like those formed
 in mechanically controlled break junctions, are shown to develop spontaneous local magnetic
order. Our density
functional calculations predict that a robust local magnetic order exists in the
atoms presenting low coordination, i. e., those forming the atom-sized
neck. 
In contrast to previous  work, we thus find that the
electronic transport can be spin-polarized, although the
net value of the conductance still agrees with available experimental
information. Experimental implications of the formation of this 
new type of nanomagnet are discussed.

\end{abstract}

 \maketitle


Fabrication of metallic nanocontacts permits to probe the electronic and
mechanical properties of conventional metals with unconventional atomic
coordination\cite{Agrait:pr:03}.  Electron transport in these systems depends 
on the tiny fraction of  atoms in the sample forming the atom-sized neck
which have a reduced coordination and are responsible for the two-terminal
resistance. Transport experiments can thereby probe the atomic and related
electronic structure of these atoms and provide information about a fundamental
question:  
How bulk properties evolve when the system reaches atomic sizes and atoms with
full bulk coordination are no longer majority. A bulk property that is
susceptible to change is magnetism.  Bulk Pt, for instance, is a
paramagnetic Fermi liquid  with a rather large spin susceptibility\cite{AHM81}. A
transition to a ferromagnetic state could be expected upon reduction of the
atomic coordination with the concomitant increase of the density of states at
the Fermi energy beyond the Stoner limit.  Density functional calculations\cite{Bahn:prl:01,Delin:prb:03,Nautiyal:prb:04}
for   one-dimensional infinite Pt chains support this hypothetical  scenario,
resulting in a ferromagnetic transition above a critical lattice spacing which,
depending on the computational approach, can be below the equilibrium lattice
constant\cite{Delin:prb:03}.  The formation of local moments in real Pt
nanocontacts would not be totally unexpected.

 Formation of and
electronic transport in finite Pt chains, for instance, have been extensively
studied experimentally\cite{Sirvent:PRB:96,Smith:PRL:01,Smith:PRL:03,Nielsen:prb:03,Rodrigues:prl:03}. 
Based upon the appearance of a peak at $G=0.5 \times 2e^2/h=0.5G_0$ in the
conductance histogram  Rodrigues {\em et al.} suggested that  Pt and Pd
nanocontacts could be spin polarized\cite{Rodrigues:prl:03}. The origin of this
peak has been later attributed to adsorbates\cite{Untiedt:prb:04} so that
magnetism in Pt and Pd nanocontacts has not been confirmed experimentally yet.
Previous theory work has
addressed the formation of local moments in Pd nanocontacts\cite{Delin:prl:04}
and in Co , Pd and Rh short chains sandwiched between Cu
planes\cite{Stepnyuk:prb:04}. To the best of our knowledge theory work on Pt
nanocontacts\cite{Vega:prb:04,Yamila:prb:04,Suarez:prb:05,Thygessen:prb:05,
jaime}  has overlooked the possibility of local magnetic order so far. Here we
perform density functional calculations of both the electronic structure and
transport  and find that local magnetic order can develop spontaneously in Pt
nanocontacts. Local magnetic moments as high as 1.2 $\mu_B$ in low-coordination
atoms are found. Interestingly,  while transport is
definitely spin polarized,  the calculated  total conductance of magnetic and
non-magnetic Pt nanocontacts is very similar and in agreement with experimental
data, explaining why magnetism  has been unnoticed so far.  

The rest of the paper is organized as follows. First we discuss the
methodology of the electronic structure calculation presented here. Second we
present results for idealized infinite and finite one-dimensional Pt chains.
Then we calculate both  zero-bias transport and the electronic structure of Pt
nanocontacts, which turn out to be magnetic beyond attainable values of the stretching. 
Finally we discuss
whether magnetism in Pt nanocontacts is robust to  thermal and quantum
fluctuations, and propose experimental verification of the different
scenarios. 
We also comment on the
effect of spin-orbit interaction, missing in our calculations.

{\em Electronic structure and transport calculations.--}  The electronic
structure of  various low dimensional structures of Pt which mimic  actual nanocontacts
are calculated in the density functional approximation,
using either CRYSTAL03\cite{CRYSTAL03} or ALACANT (ALicante Ab initio Computation Applied to Nanotransport), our \textit{ab initio}
transport package\cite{Palacios:prb:01} interfaced to
GAUSSIAN03\cite{Gaussian:03}. ALACANT describes the bulk electrodes using a
semiempirical tight-binding Hamiltonian on a Bethe lattice. The
spin-resolved density matrix includes the effect of the electrodes by
means of self-energies and the spin-polarized transport\cite{Palacios:prb:05,Jacob:prb:05} is
obtained with the standard Landauer formula.  Both
GAUSSIAN03/ALACANT and CRYSTAL03 perform electronic structure
calculations using a basis of  localized atomic orbitals (LAO). CRYSTAL03
permits to calculate infinite systems with crystalline symmetry whereas
ALACANT is suitable for systems with no symmetry or periodicity such as nanocontacts in or 
out-of-equilibrium situations. In the case of the Pt
calculations shown here we use scalar relativistic (SR) pseudopotentials for the 60 inner electrons and
the remaining 18 electrons are treated using generalized gradient approximation (GGA) density
functionals. The basis set  used for all the calculations has been optimized to
describe bulk Pt as well as Pt surfaces\cite{Doll:SS:04}. Other basis sets such as
LANL2DZ or SDD\cite{Gaussian:03} have occasionally been employed for
comparison.  The main results do not depend on the choice of  basis set.

\begin{figure}
[t]
\includegraphics[width=3.in,height=2.7in]{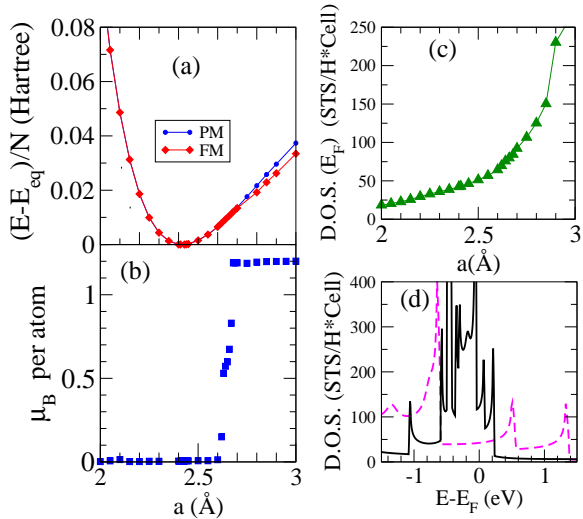}
\caption{ \label{fig1}(Color online).
(a) Energy per atom for a perfect monostrand Pt chain. (b) Magnetic moment per atom as a
function of lattice spacing $a$. (c) Density of states (D.O.S.) of the paramagnetic
chain at the Fermi energy
 as a function of $a$.  (d) D.O.S. as a function of energy for 
 $a=2.4\AA$ (dashed) and $a= 2.8\AA$ (solid) } 
\label{chain}
\end{figure}

\textit{Results.--} We first consider a perfect one-dimensional mono-strand Pt
chain. Such an idealized system serves as a standard starting point to
understand lower symmetry geometries. It also permits to  test whether our
LAO pseudopotential methodology  reproduces the
results obtained with SR all electron plane-wave calculations
reported  by Delin {\em et al.}\cite{Delin:prb:03}. In Fig. \ref{chain}(a) we
show the energy per atom as a function of the lattice constant $a$ both for
the paramagnetic (PM) and the ferromagnetic (FM) chain. 
They both have a minimum at $a=2.4$\AA. The FM chain 
develops a non-negligible magnetic moment when the lattice constant goes
beyond  $a\simeq 2.6$\AA. This configuration is clearly lower in energy above
that distance. The energy difference between the FM and the PM configurations
is  16 meV per atom for $a=2.7\AA$  and 33 meV per atom for $a=2.8$\AA.  The 
magnetic moment per atom reaches a saturation value of $1.2\mu_B$.    The
equilibrium distance, critical spacing, asymptotic magnetic moment and shape of
the phase boundary obtained by us are similar to those obtained by Delin {\em et
al.}\cite{Delin:prb:03} using a SR all-electron plane-wave calculation.   Our results and those of Delin
et al. underestimate the onset of the
magnetic  transition compared to calculations including spin-orbit coupling\cite{Delin:prb:03,Nautiyal:prb:04} that predict that a magnetic moment forms
already below the equilibrium distance.

The magnetic transition in the phase diagram [Fig. 1(b)]  is compatible with
the Stoner criterion for ferromagnetic instability.  As the chain is stretched,
the atom-atom coupling becomes weaker, the bands narrow down and so does the
density of states ($\rho(\epsilon)$).  Since the integrated $\rho(\epsilon)$
must be equal to the number of electrons per atom, narrowing of the DOS implies
an increase of the $\rho(\epsilon)$ [see Fig. \ref{chain}(d)] and, therefore,
an increase of the spin susceptibility, which is proportional to
$\rho(\epsilon_F)$.  In Fig. \ref{chain}(c) we show how the $\rho(\epsilon_F)$
of the PM chain increases as a function of the lattice constant. The remarkable
feature of  Pt chains is that the Stoner instability occurs close to the
equilibrium lattice spacing.

\begin{figure}
[t]
\includegraphics[width=3.in,height=2.5in]{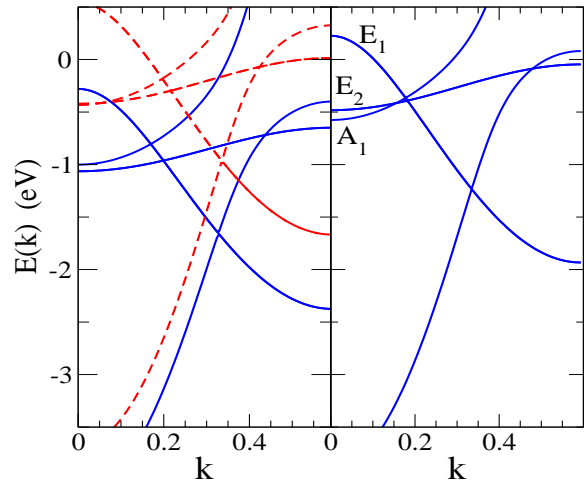}
\caption{ (Color online).
Energy bands for ideal Pt chain with $a=2.8\AA$. Left: ferromagnetic phase.
Right: paramagnetic case.  } 
\label{fig2}
\end{figure}

The electronic structure of the ideal Pt chain sheds some light on the
electronic structure of the nanocontact.
In Fig. \ref{fig2} we show the energy bands for the ideal Pt chain both in the
FM (left panel) and PM (right panel) configurations, 
for a lattice spacing $a=2.8\AA$. We notice that the spectrum at $k=0$
has four resolved energy levels per spin. These correspond to the $6s$ level, and
the 5$d$ levels which, because of the axial potential created by the
neighboring atoms,  split into 2 doublets $E_1$, $E_2$ and one singlet
$A_1$. The $E_1$ and $E_2$ are linear combinations  of  orbitals with 
$L_z=\pm 1$ and $L_z=\pm 2$,
respectively, whereas the $A_1$ singlet is a $L_z=0$ orbital that hybridizes
with the lower energy $6s$ orbital. The largest contribution to the density of
states, and therefore to the magnetic instability, comes from the $A_1$-like band at the edge
of the Brillouin zone.  However, the prominent
role played by these bands in the magnetic behavior of Pt chains is in stark contrast 
with their role on the transport properties of Pt nanocontacts 
(see below and see also related work on Ni nanocontacts\cite{Jacob:prb:05}). 
Four spin-degenerate bands cross the Fermi energy in the
PM case whereas  7 spin-split bands do it in the FM chain.  In the
FM chains the number of spin minority channels is 6, and the
number of spin majority channels is 1. Although spin-orbit interaction modifies 
significantly the bands \cite{Delin:prb:03},
the number of bands at the Fermi energy is pretty similar in both cases. 
Therefore, one can anticipate that 
the number of open channels in the magnetic and
non-magnetic Pt nanocontacts studied below should be roughly
the same and thereby the conductance should be similar, 
although the spin polarization might well be large in the former case. 
The conductance of the ideal FM chain is $3.5 G_0$, very far
from the value of  0.5$G_0$ that allegedly signals the emergence of magnetism
and also
far away from half of the conductance of the PM chain, so it is very
unlikely that the celebrated half quantum can be attributed exclusively 
to magnetism.

Real  Pt chains are typically less than five atoms long and are connected to
bulk electrodes. Although not surprising, we have verified that magnetism
survives in isolated short chains with $N_A=$ 3, 4 and 5 Pt atoms.  
The equilibrium distance is 2.4\r{A} for all  $N_A=3$, $N_A=4$, and $N_A=5$.
Interestingly, the short chains are always magnetic in the $N_A=3$ and $N_A=4$
cases and show a non-magnetic to magnetic crossover at $a=2.6$\r{A} in the
$N_A=5$ case, already similar to the ideal infinite chain. The total  magnetic moment 
of all the $N_A=3$ chains with $a<3.0 $\r{A} is 4$\mu_{\rm B}$. The outer atoms have a
magnetic moment of $1.36\mu_{\rm B}$ and the central atom with larger coordination has a
smaller magnetic moment of $1.29\mu_{\rm B}$. In the case of $N_A=4$ the total 
magnetic moment is 6$\mu_B$, and their distribution is similar to the $N_A=3$ case.

\begin{figure}
  \includegraphics[width=\linewidth]{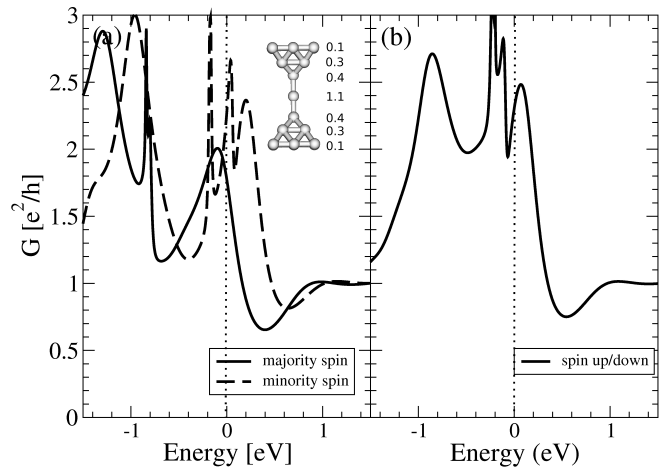}
  \caption{ 
    Conductance per spin channel for a nanocontact with a 
    3-atom Pt chain (see inset) for the magnetic solution (a) 
    and the non-magnetic one (b). 
    The atom-atom distance in the chain is $2.8$\r{A}.} 
  \label{Pt-nanocontact} 
\end{figure}

The calculations above show that magnetism is present both in finite- and
infinite-sized Pt systems with small atomic coordination. It remains to be seen
that this holds true in nanocontacts where none or only few atoms have a small
coordination (like in the case of formation of short chains),  but these are strongly
coupled  to the bulk.  In order to verify  whether or not this is the case, we
have calculated  both electronic structure and transport for a model Pt
nanocontact. It is formed by two opposite pyramids grown in the (001)
crystallographic orientation of bulk Pt and joined by one atom which presents
the lowest possible coordination [see inset in Fig. \ref{Pt-nanocontact}(a)].
Relaxation of the 11 inner atoms of the cluster has been  performed starting
from an equilibrium situation as a function of the distance $d$ of the outer
planes. Zig-zag configurations appear in the chain for small values of $d$ (not
shown)\cite{jaime} until the three-atom chain straightens up (see inset in Fig.
\ref{Pt-nanocontact}(a)] followed by a plastic deformation (not shown). This
deformation can be in the form of a rupture or a precursor of the addition of a
new atom to the chain which comes from one of the two 4-atom
bases\cite{Bahn:prl:01}. 

We now compute the transmission before the plastic deformation occurs (left
panel in Fig. \ref{Pt-nanocontact}), where the atom-atom  distance in the short
chain  is 2.82 \AA.  The value of the corresponding
atomic-plane-averaged  magnetic moments are also shown in the inset. As
expected, it decreases for atoms in the bulk as the coordination reaches the
bulk value.
Some atomic realizations in the stretching process (like zig-zag ones) 
result in nanocontacts with smaller
Pt-Pt distance and no magnetism, in agreement with the infinite chain phase
diagram in Fig. 1.  
In contrast to the infinite chain, there are only  three channels contributing
to the total conductance for minority electrons and there are more than one
(three) for  majority ones. For the majority electrons these are a perfectly
transmitting $s$-type channel and two partially open ($T=0.4$) $pd$-type
channels (one $p_xd_{xz}$- and one $p_yd_{yz}$- hybridized). The three minority
channels have the same character as the majority channels, except that  here
the $s$-type  channel does not transmit perfectly while the transmission of two
$pd$-type channels is enhanced so that all three minority channels have a
transmission around $0.7$. The other two remaining $pd$-like channels are
responsible for the sharp resonances that appear around the Fermi level. The
total conductance of the nanocontact in the FM case thus turns out
to be around $4 e^2/h = 2 G_0$  which is only slightly larger than the average
experimental value corresponding to the last plateau  (1.75$G_0$), but,
interestingly, barely differs from the value obtained when the possibility of
magnetic  order is ignored [around $2.3 G_0$, see right panel in Fig.
\ref{Pt-nanocontact}].  As in the case of the ideal chain, the FM
conductance is not half  of the PM conductance nor half of $G_0$. To
conclude this discussion we notice the, although transport is only weakly spin
polarized, magnetism brings the $pd$-like resonances up to the Fermi level
compared to the non-magnetic case. These resonances may well give features in
the low bias conductance not present if Pt were not magnetic.

{\em Conclusions and Discussion.--} 
The main conclusion of this work is that density functional calculations predict that 
nanochains formed in Pt nanocontacts can be stretched as to become magnetic.
The magnetic moment is localized mainly in the atoms with small coordination and
does not modify appreciably the total conductance, although the transmission is moderately
spin polarized.  How robust are these results? 
It is well known that both   local and gradient-corrected density functionals 
present some degree of  electronic self-interaction, in contrast with the
Hartree-Fock approximation. Self-interaction is larger for localized electrons
and shifts the  the $d$ bands upwards in energy, as shown  in the case of Co,
Ni and Pd one dimensional chains\cite{Wierzbowska:prb:05}. A number of schemes
to avoid this problem, like LDA+U  and Self-Interaction Correction functionals
have been proposed.  The method of choice between chemist is hybrid
functionals\cite{Becke:JCP:93} in which local and Hartree-Fock exchanged are
combined and the self-interaction is reduced.   We have calculated  the
magnetic phase diagram of the one dimensional  Pt chain using the hybrid B3LYP
functional\cite{Becke:JCP:93} and found, somewhat expectedly, that magnetism is
enhanced and that B3LYP infinite Pt chains are ferromagnetic down to the
equilibrium distance ($a=2.4$ \AA). Both non-local exchange and 
spin-orbit coupling \cite{Delin:prb:03} enhance the stability of 
magnetism in Pt nanocontacts.  This and previous results on Ni
nanocontacts\cite{Jacob:prb:05} lead us to believe that self-interaction is an
issue in the electronic structure and transport properties of transition metal
nanocontacts and further work is necessary along these
lines\cite{Ferretti:prl:05}.

The  mean field picture of the electronic structure describes a static  magnetic moment 
without preferred spatial direction. 
In reality the nanomagnet formed in the break junction is
 exchanged coupled dynamically to the Fermi
sea of the conduction electrons of the electrodes. In the case of a spin
$S=1/2$ this can result in the formation of a Kondo singlet that would yield
an anomaly in the zero bias conductance.  For larger spins, the
conduction electron sea cannot screen the spin completely so that the magnetic
moment survives. The magnetic moment of the 
nanocontact in Fig. 3 is  $S\approx 6$,
comparable to that of single molecule magnets
\cite{Mn12}, making the formation of a Kondo singlet unlikely. 

In the absence of spin-orbit interactions and external magnetic field a
electronic configuration with total spin $S$ has $2S+1$ degenerate
configurations corresponding to the spin pointing along different directions.  
However, spin-orbit interaction is strong in Pt and produces  spin anisotropy, 
favoring orientation along the  transport direction axis in the case of one
dimensional chains\cite{Delin:prb:03}.  Thermal fluctuations of the magnetic
moment between these two configurations are quenched for  temperatures smaller
than the  anisotropy  barrier.  Departures from the easy axis orientation
will be  damped via electron-hole pair creation across the Fermi energy that
would also result in  small bias features in transport\cite{Carlos}. The
application of a sufficiently strong magnetic field in the direction
perpendicular to the easy axis moves the  local magnetic moments away from
their easy axis. This is known to change the number of open channels at the
Fermi energy in both the case of Ni ideal chains\cite{Velev:prl:05} and in the
case of ferromagnetic semiconductor tunnel junctions \cite{Brey:apl:04}.  This
effect,  or maybe even larger, can be expected in Pt nanocontacts and  could be
used to detect the nanomagnetism experimentally.

Fruitful discussions with J. Ferrer are acknowledged. Work funded from  Grants No.
MAT2002-04429, FIS2004-02356, the Ramon y Cajal Program  (MCyT, Spain),
and  Grant No. GV05/152 from Generalidad Valencina,   are acknowledged. This work 
has been also partly funded by FEDER funds.


\widetext
\end{document}